\documentclass[a4paper,fleqn,usenatbib]{mnras}

\usepackage[T1]{fontenc}
\usepackage{ae,aecompl}

\usepackage{graphicx}	
\usepackage{amsmath}	
\usepackage{amssymb}	
\usepackage{multirow}
\usepackage{xcolor}

\newcommand{\maxi}{\textit{MAXI}}
\newcommand{\nicer}{\textit{NICER}}
\newcommand{\nustar}{\textit{NuSTAR}}

\newcommand{\vsgr}{V4641~Sgr}
\newcommand{\vcyg}{V404~Cyg}
\newcommand{\eighteen}{Swift~J1858.6--0814}

\title[Type I bursts in Swift J1858.6--0814]{Discovery of thermonuclear (Type I) X-ray bursts in the X-ray binary Swift~J1858.6--0814 observed with \nicer\ and \nustar}

\author[D. J. K. Buisson et al.]{D. J. K. Buisson$^{1}$,\thanks{Email: d.j.k.buisson@soton.ac.uk}
  D. Altamirano$^{1}$,
  P. Bult$^{2,3}$,
  G. C. Mancuso$^{4,5}$,
  T. G\"{u}ver$^{6,7}$,
\newauthor   G. K. Jaisawal$^{8}$,
  J. Hare$^{3}$\thanks{NASA Postdoctoral Fellow},
  A. C. Albayati$^{1}$,
  Z. Arzoumanian$^{3}$,
  N. Castro Segura$^{1}$,
\newauthor  D. Chakrabarty$^{9}$,
  P. Gandhi$^{1}$,
  S. Guillot$^{10}$,
  J. Homan$^{11,12}$,
  K. C. Gendreau$^{3}$,
\newauthor  J. Jiang$^{13,14}$,
  C.~Malacaria$^{15,16}$\thanks{NASA Postdoctoral Fellow}, 
  J. M. Miller$^{17}$,
  M. \"{O}zbey Arabac\i$^{1,18}$,
\newauthor  R. Remillard$^{9}$,
  T. E. Strohmayer$^{19}$,
  F. Tombesi$^{2,20,3,21}$,
  J. A. Tomsick$^{22}$,
\newauthor  F. M. Vincentelli$^{1}$ and
  D. J. Walton$^{23}$\\
  Affiliations at end.
}

\date{Accepted 2020 September 7. Received 2020 September 7 in original form 2020 June 12}

\pubyear{2020}

\begin{document}
\label{firstpage}
\pagerange{\pageref{firstpage}--\pageref{lastpage}}
\maketitle

\begin{abstract}
\eighteen\ is a recently discovered X-ray binary notable for extremely strong variability (by factors $>100$ in soft X-rays) in its discovery state. We present the detection of five thermonuclear (Type I) X-ray bursts from \eighteen, implying that the compact object in the system is a neutron star.
Some of the bursts show photospheric radius expansion, so their peak flux can be used to estimate the distance to the system. The peak luminosity, and hence distance, can depend on several system parameters; for the most likely values, a high inclination and a helium atmosphere, $D=12.8_{-0.6}^{+0.8}$\,kpc, although systematic effects allow a conservative range of 9-18\,kpc.
Before one burst, we detect a QPO at $9.6\pm0.5$ mHz with a fractional rms amplitude of $2.2\pm0.2$\% ($0.5-10$\,keV), likely due to marginally stable burning of helium; similar oscillations may be present before the other bursts but the light curves are not long enough to allow their detection.
We also search for burst oscillations but do not detect any, with an upper limit in the best case of 15\% fractional amplitude (over $1-8$\,keV).
Finally, we discuss the implications of the neutron star accretor and this distance on other inferences which have been made about the system.
In particular, we find that \eighteen\ was observed at super-Eddington luminosities at least during bright flares during the variable stage of its outburst.
\end{abstract}

\section{Introduction}

A key aspect of accreting systems is the object onto which the accretion is occurring; in X-ray binaries (XRBs) this is either a neutron star (NS) or black hole (BH).
Many observable properties are similar in either case, so determining which is present is often a challenging task.

There are several properties which can divide NSs and BHs as populations and some features which empirically appear to occur in only one type of system.
Firstly, outbursts of the different classes of source follow different tracks in gross properties such as the hardness-intensity or colour-colour diagrams \citep[e.g.][]{vanderklis06}. However, this requires monitoring of the full outburst and some sources do not follow the typical patterns.
Additionally, quasi-periodic oscillations (QPOs) are only found at kHz frequencies in neutron star systems \citep{vanderklis96,strohmayer96}, although there is not yet a universally accepted model for their production \citep[e.g. review by][]{vanderklis06}.

Also, BH and NS systems can be separated in the radio/X-ray luminosity plane (while in the hard state),  with  BH systems being radio brighter \citep{migliari06,gallo18}.
Similarly, the hard Comptonised component tends to have a higher temperature in BH systems \citep{burke17}.
However, the loci of BHs and NSs overlap in these properties, so they cannot be used to determine the accretor definitively in an individual source, particularly where a source shows unusual properties.

Other properties of an accreting system can give a definitive determination of whether the accreting object is a black hole or neutron star.
To confirm a black hole accretor requires a dynamical mass measurement which is greater than possible neutron star masses \citep[e.g.][]{webster72,bolton72,orosz97}, since there are no particular accretion properties which are unique to black holes. Conversely, there are several properties which are confirmed as unique to neutron stars, since the neutron star surface can provide an additional location for emission components and they can support large scale magnetic fields.
The emission from this surface may be detected directly as a soft ($0.1-0.3$\,keV) blackbody-like component \citep[e.g.][]{brown98}. This component is much fainter than the accretion luminosity, so cannot be identified during the first outburst in which a source is detected and requires sensitive observations to detect.
Also, neutron stars can pulse coherently on their spin period, which can be observed at wavelengths from radio \citep{hewish68} to X-ray \citep[e.g. review by][]{patruno12}.
A further feature of X-ray binaries particular to those hosting neutron stars is Type I X-ray bursts \citep[e.g.][]{grindlay76,hoffman78,lewin93,strohmayer06,galloway08rxte,galloway17}.

Type I X-ray bursts occur due to explosive thermonuclear burning of accreted material on the neutron star surface.
As material accretes onto the neutron star, it adds to layers of hydrogen and helium on the surface. When the pressure at the bases of these layers becomes large enough, it will ignite thermonuclear burning. Depending on the conditions, this burning may occur stably, contributing to the power in the persistent emission, or explosively, producing Type I bursts.
These bursts can be ignited by hydrogen and/or helium (or occasionally carbon in `superbursts', e.g. \citealt{cornelisse00,cumming01,strohmayer02,intzand17}), depending on the stability of each burning process, which depends principally on the accretion rate \citep{narayan03} but also the metallicity of the accreted material and the internal temperature of the neutron star \citep{bildsten95,bildsten98,cumming04}.
The basic classes of burning are thought to be as follows; while these classes of burning are reproduced in most numerical studies of Type I bursts, the exact values of accretion rate at which each one occurs differ between works \citep[e.g.][]{fujimoto81,bildsten98,narayan03}.

At low accretion rates, a layer of hydrogen builds up before it reaches sufficient pressure to begin burning. At this point, the energy released by ignition causes most of the hydrogen layer to burn rapidly,  which is observed as a Type I burst.
At higher accretion rates, the accreted hydrogen promptly reaches sufficient temperature and pressure to burn so the burning is stable.
When sufficient helium builds up, the pressure ignites helium burning which is responsible for the burst.
At still higher accretion rates, not all of the hydrogen can be burnt before helium ignition occurs; this hydrogen is burnt along with the helium in a mixed burst.
At the highest accretion rates, both hydrogen and helium burning occur stably so no Type I bursts are observed.
However, close to the transition to stability, the burning is marginally stable and has an oscillatory mode \citep{heger07qpo}, which has been used to explain the millihertz QPOs observed prior to some Type I bursts \citep{revnivtsev01,altamirano08,lyu16,mancuso19}.

There are also other effects which can affect the occurrence and type (fuel) of thermonuclear bursts. 
Each burst may not burn all of the available fuel, so some hydrogen and helium will remain after one burst and can affect properties of following bursts. Similarly, the burnt material will contain additional carbon, nitrogen and oxygen from helium burning. These nuclei catalyse hydrogen burning so can affect later bursts as well.
The neutron star spin \citep{spitkovsky02,galloway18} and the geometry of where on the star the material is accreted \citep{kajava14} can also affect burst properties.

Type I bursts can sometimes be used as standard candles, as they can be bright enough to reach the Eddington limit. In this situation, the radiation pressure lifts material in the NS surface and the atmosphere expands in Photospheric Radius Expansion \citep[PRE;][]{tawara84,lewin84}.
Since the Eddington limit is only weakly dependent on radius, this produces a period during which the luminosity remains constant at the (known) Eddington value. This PRE phase may be identified (and distinguished from a simple plateau in the burning rate) by measuring the change in photospheric radius from the time-resolved X-ray spectrum.
The measured flux during the PRE phase may then be used with this standard candle to estimate the distance to the source \citep{vanparadijs78,kuulkers03}.

\subsection{\eighteen}

\begin{figure}
\includegraphics[width=\columnwidth]{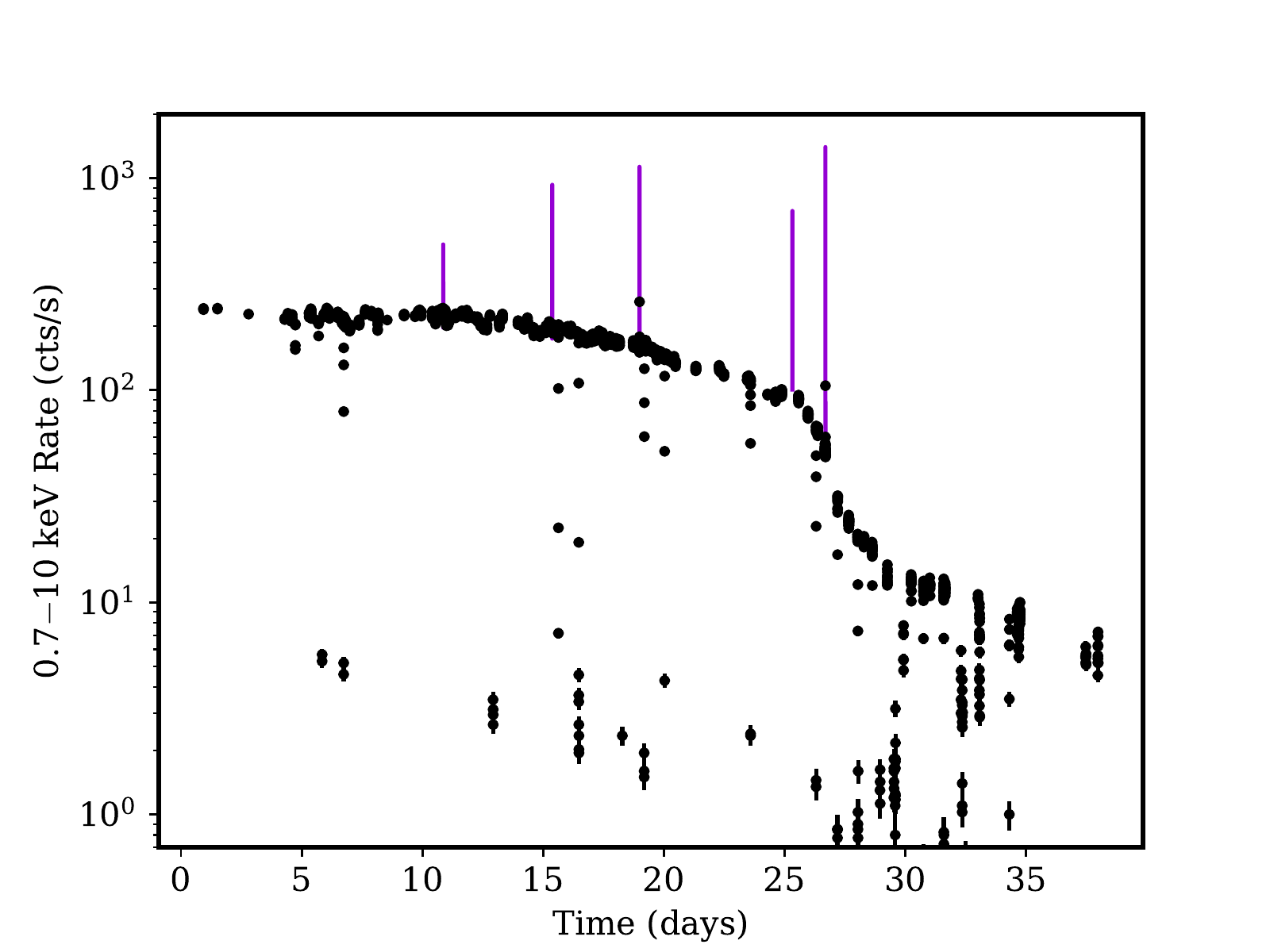}
\caption{\nicer\ light curve of \eighteen\ since leaving Sun constraint in 2020, showing times of observed Type I bursts (purple). In addition to the long-term flux decrease, several dips and eclipses are visible; these will be considered in detail in future work. The full \nicer\ light curve is shown in black at a resolution of 40\,s and the bursts (purple) extend to their maximum count rate at 0.1\,s resolution. The zero-point for the time axis is the start of 2020\,February\,25 (MJD 58904).}
\label{fig:lc20}
\end{figure}

The low-mass X-ray binary \eighteen\ has been in its first observed outburst since late 2018 \citep{krimm18}.
The X-ray emission in the initial phase of the outburst was highly variable as was the emission in other wavebands \citep[Fogantini et al. in prep.]{ludlam18_1858,vandeneijnden20}: the \nicer\ $0.5-10$\,keV count rate peaks at over 650\,cts/s within 200\,s of intervals at $\approx2.5$\,cts/s (Fogantini et al. in prep.), much larger than the typical tens of percent RMS on these timescales \citep{mcclintock06}. We refer to this stage of the outburst (all observations in 2018 and 2019) as the flaring state.
The X-ray spectra were also extremely hard: $\Gamma<1$ if fitted with a simple powerlaw, \citep{kennea18,ludlam18_1858}, compared to typical $\Gamma>1.5$ \citep[e.g.][]{zdziarski99}. This may be explained by the contribution of reflection and absorption: they also show a strong neutral iron K$\alpha$ line and K edge \citep{reynolds18,hare20} and soft X-ray emission lines \citep{buisson20rgs}.
It also shows P-Cygni lines in its optical spectra, which look similar to those seen in several BH XRBs \citep[Castro-Segura in prep.]{munoz19,munoz20}, as well as strongly variable optical emission \citep{paice18}.
These properties have led to \eighteen\ being viewed \citep{hare20} as an analogue of \vcyg\ \citep{gandhi16,walton17,motta17abs} and \vsgr\ \citep{wijnands00,revnivtsev02}, which have been dynamically confirmed as hosting black holes \citep[][respectively]{casares92,orosz01}. 
\eighteen\ also lies within the range occupied by BHs in the radio-X-ray plane \citep{vandeneijnden20}.
However, recent observations of \eighteen\ have shown qualitatively different X-ray properties, suggesting a state change while the source was unobservable due to Sun constraint (between 2019 November and 2020 February), although the properties of the initial phase were not typical of a canonical state \citep[e.g.][]{vanderklis94sim}. In the 2020 observations, the flux level is much steadier and the strong iron line and edge are absent \citep[and Figure~\ref{fig:lc20}]{buisson20atel1}, . These observations have also shown Type I X-ray bursts in both \nicer\ and \nustar\ data \citep{buisson20atel2}, unambiguously identifying the compact object as a neutron star.

In this paper, we analyse the Type I X-ray bursts detected in \nicer\ and \nustar\ data of \eighteen.

\section{Observations and data reduction}
\label{sec:odr}

We have inspected the \nicer\ \citep{gendreau16} light curves from 2020 by eye. Apparent Type I bursts are present in OBSIDs 3200400106, 3200400111, 3200400114, 3200400121 and 3200400122, corresponding to March 6, 11, 14, 21 and 22.

We begin with the calibrated, unfiltered events file from HEASARC (\texttt{event\_cl/ni32004001**\_0mpu7\_ufa.evt}).
We use the standard filters\footnote{For further information on the filters, see heasarc.gsfc.nasa.gov/lheasoft/ftools/headas/nimaketime.html} to produce good time intervals (GTIs) apart from the undershoot range, which we relax from $\leq200$\,s$^{-1}$ to $\leq300$\,s$^{-1}$ for the first Type I burst and $\leq250$\,s$^{-1}$ for the second. This is required due to high optical loading due to the relatively low Sun angle.
Additionally, to include the peak of the second burst, we relax the offset from the nominal target direction slightly, using 0.0155$^\circ$ rather than 0.015$^\circ$.
This is a small change from the standard value, so data during this time are unlikely to show significant deviations from the standard calibration.
Further, the fourth  burst occurs during passage through the South Atlantic Anomaly (SAA) and the overshoot rate reaches close to 5\,s$^{-1}$, so is removed by standard filtering. We remove these filters in order to show the light curve but note that the spectrum may be affected.

We then use \textsc{nicerclean} to produce a clean events list, which we then barycentre to the ICRS reference frame and JPL-DE200 ephemeris. From this, we extract spectra and light curves using \textsc{xselect}.

We use \nustar\ \citep{harrison13} OBSID 90601308002, which overlaps with \nicer\ OBSID 3200400106. We reduce this using the standard \textsc{nupipeline} and \textsc{nuproducts} software, version 1.9.0.
We use a source region of a circle of radius 2\,arcmin centred on the centroid of the detected counts. We use a background region of a circle of radius 2\,arcmin from a source-free area of the detector.

\section{Results and Discussion}
\label{sec:res}

\subsection{Long-term light curve and burst recurrence time}

We show the light curve of \eighteen\ since leaving Sun constraint on 2020 February 25 in Figure~\ref{fig:lc20}. 
The count rate shows a secular decrease throughout the whole of this period, punctuated by short dips and eclipses as well as the five Type I bursts analysed here.	The drop in persistent count rate from the first to last burst was by a factor of around 4 and bursts were in general brighter at lower persistent count rate, with only the fourth burst not following this trend. Around the time of the last observed burst, the rate started decreasing more rapidly, before flattening once more. The period of fastest flux drop extended considerably before and after the final burst, so the coincidence in time is probably only by chance.
As well as the Type I bursts, several dips are present, many of which are due to eclipses \citep{buisson20atel2}; these will be analysed in detail in future work. 

The times between Type I bursts are 4.5, 3.6, 6.3 and 1.4\,days (we summarise lists, such as this, of properties of each burst in Table~\ref{tab:summ}). Since the coverage of \eighteen\ is not continuous, there may have been other bursts between those observed, in observation gaps. Therefore, these gaps are an upper limit to the recurrence time. The duty cycle of \nicer\ observations is low ($\approx3.9$\% over the 37\,days shown in Figure~\ref{fig:lc20}, but not evenly across this time) so it is very likely that other bursts did occur outside times of observation.
Furthermore, we can consider the $\alpha$-value, the ratio of inter-burst (persistent) fluence to burst fluence, which is typically $\approx40$ for hydrogen fuelled bursts and $\approx100-200$ for helium \citep[e.g.][]{gottwald86,galloway04}. Here, the lowest observed $\alpha\approx500$ (for burst~5, integrating the fluxes found in Section~\ref{sec:spec}) is higher, meaning more emission occurs between bursts than would be expected. This suggests that other, intermediate bursts did occur and/or substantial nuclear burning occurred between bursts.

There is also a period longer than any gap between observed bursts at the start of the \nicer\ monitoring ($\approx10$\,days) where no bursts are observed; again, it is possible that bursts did occur during this period but that they occurred during gaps in the \nicer\ monitoring (which observed only 0.24\,days of this time). The count rate and spectral shape show no large changes during this time, so there is no obvious reason for bursts not to have occurred.
An alternative explanation for the lack of bursts in this period is that it followed a superburst, which quenched the normal Type I bursts \citep{keek12sb}; however, there is no evidence in the \nicer\ monitoring or \maxi\ data (which cover earlier times) for a superburst having occurred.

\subsection{Confirmation of source of Type I bursts}

\begin{figure*}
\includegraphics[width=\columnwidth]{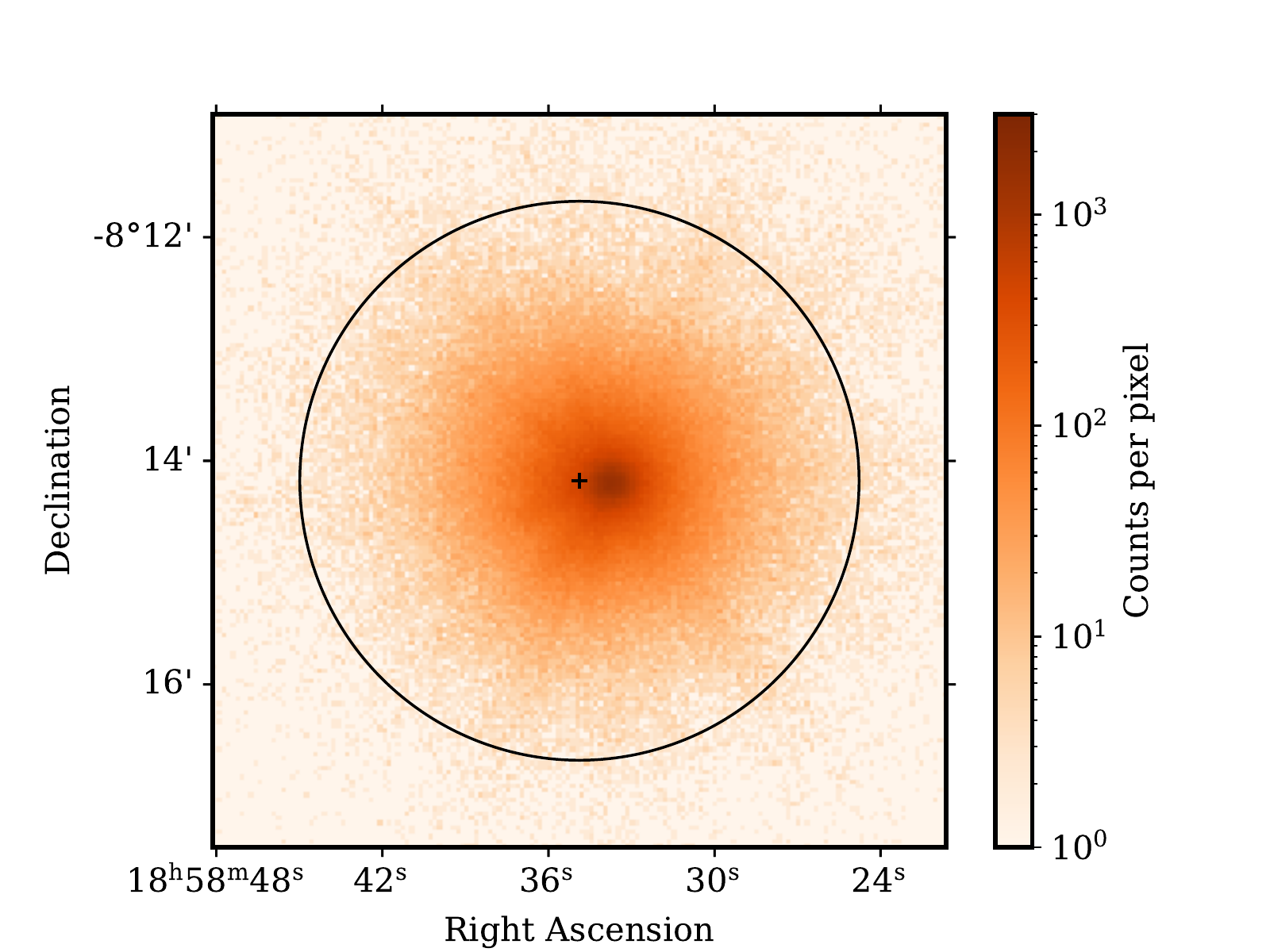}
\includegraphics[width=\columnwidth]{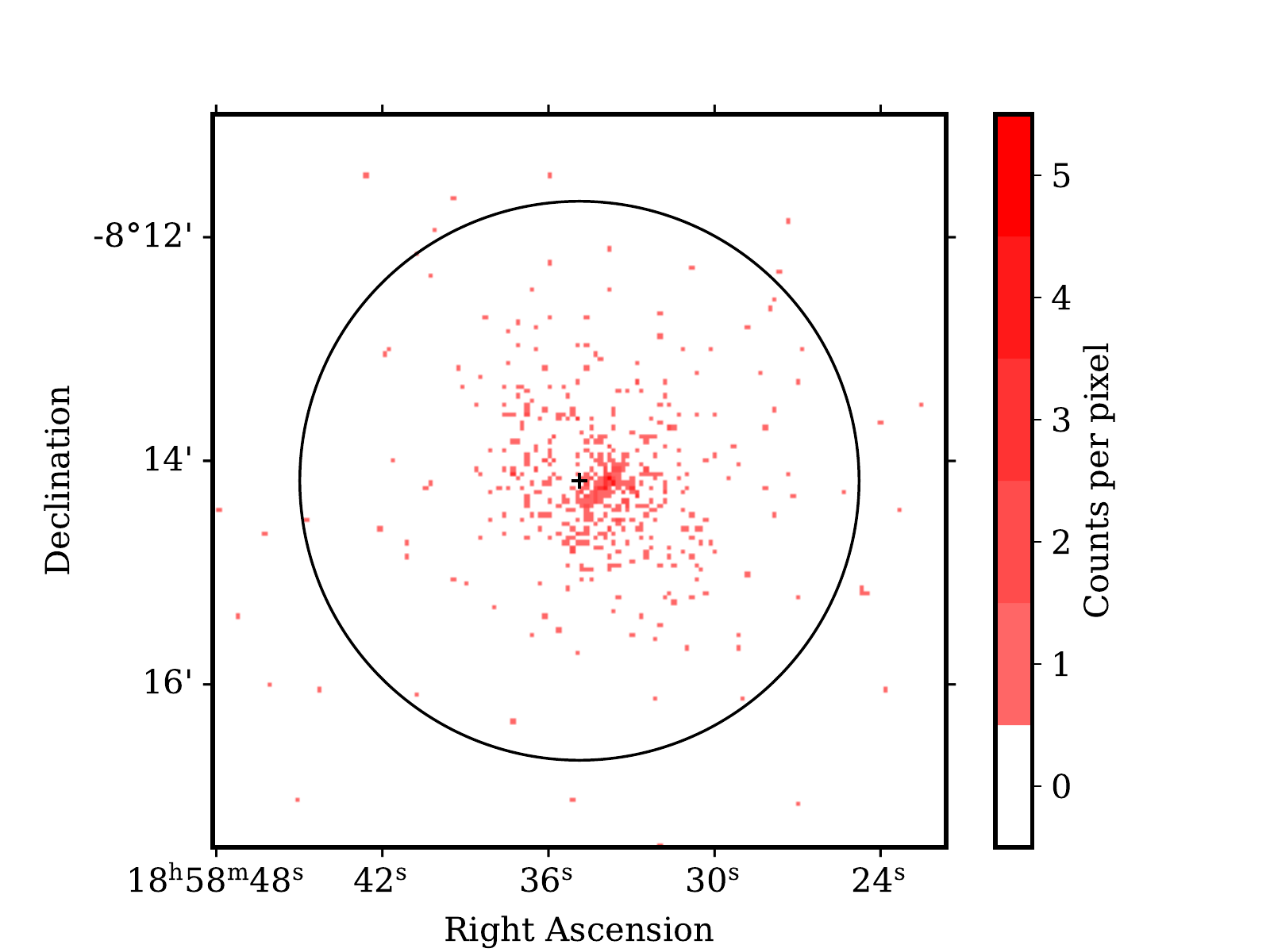}
\caption{$3-50$\,keV \nustar\ image of the sky around \eighteen. Left: over the full (26.7\,ks on source time) observation; only a single point source is apparent, at the position of \eighteen. Right: during the Type I burst only; the source position matches the position during the full observation.  The \nicer\ field of view is shown by the black circle and the nominal pointing direction by the black cross.}
\label{fig:im}
\end{figure*}

The first burst was observed by both \nustar\ and \nicer.
We show a \nustar\ image of the sky around \eighteen\ in Figure~\ref{fig:im}. This shows that, to the resolution available to \nustar, only one source is apparent in the \nicer\ field of view and the location of the Type I burst flux is consistent with the location of the persistent emission. The offset between the \nustar\ position and the nominal \nicer\ pointing is around 15\,arcsec, which is less than the 1\,arcmin nominal pointing stability of \nicer\ \citep{arzoumanian14}.
This shows that the X-ray bursts are from \eighteen.

\subsection{Type I X-ray burst light curves}

\begin{figure*}
\includegraphics[width=\textwidth]{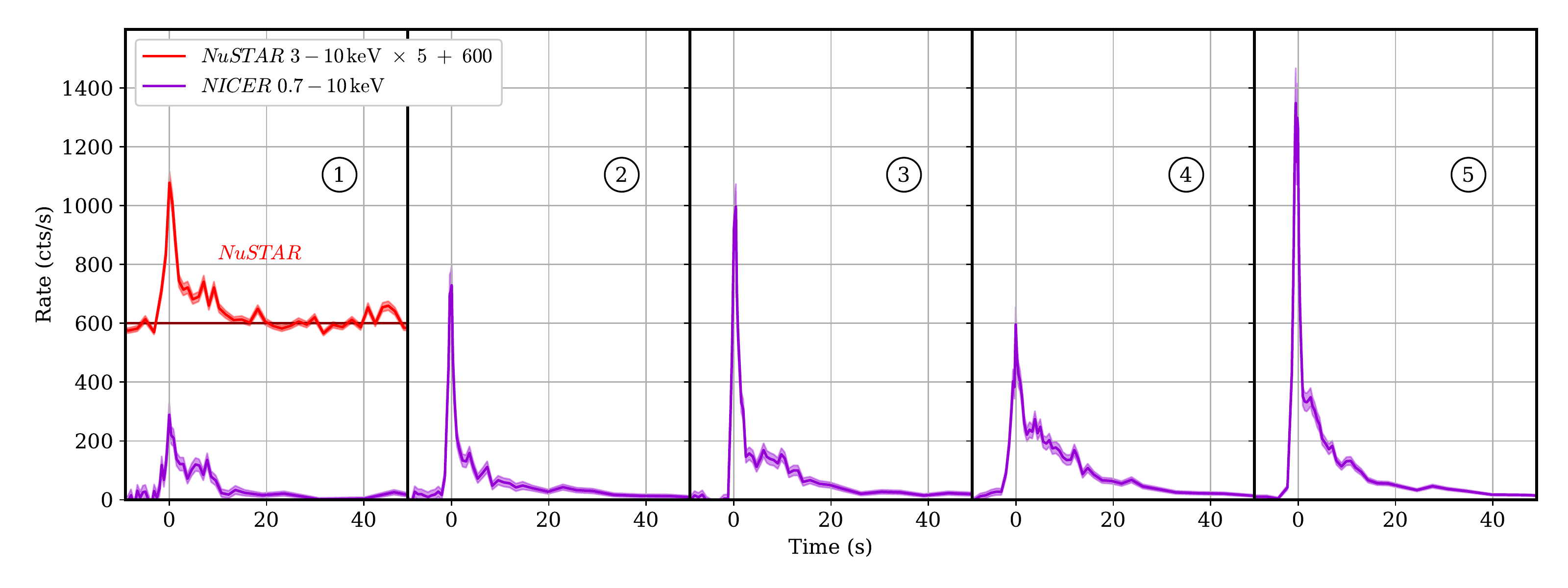}
\caption{Light curves of each Type I burst, in order of occurrence. Purple: $0.7-10$\,keV \nicer; red: $3-10$\,keV \nustar, scaled (increased by a factor of $5$) and offset (by $+600$\,cts\,s$^{-1}$). 
Each burst has had the persistent rate (the mean rate from $50-200$\,s before the burst) subtracted. The shaded regions are the $1\sigma$ Poisson uncertainties.
}
\label{fig:lcs}
\end{figure*}

\begin{figure}
\includegraphics[width=\columnwidth]{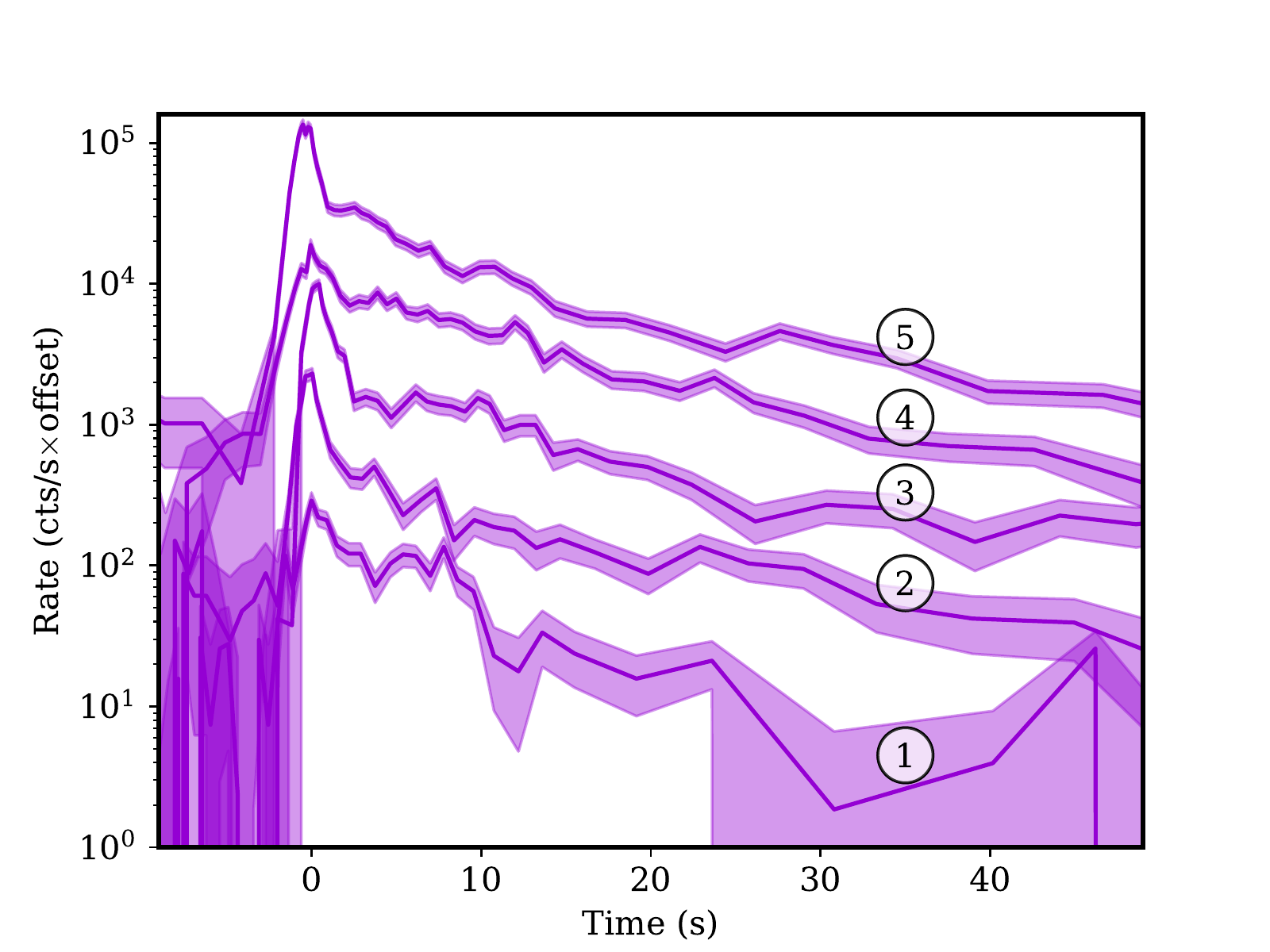}
\caption{Light curves of each Type I burst, in order of occurrence from bottom to top. Purple: $0.7-10$\,keV \nicer. Each burst is offset from the previous by a factor of $10^{0.5}$.
Each burst has had the persistent rate (the mean rate from $50-200$\,s before the burst) subtracted.
The cooling tail is similar for each burst but bursts 2, 3 and 5 have a stronger initial peak. The shaded regions are the $1\sigma$ Poisson uncertainties.
}
\label{fig:lc_stack}
\end{figure}

The light curves for each Type I burst are shown in Figure~\ref{fig:lcs}. 
Each burst has a fast rise, lasting $\lesssim3$\,s, a single peak and fades to being undetectable over the persistent level within up to $\approx40$\,s.
The decay of each burst, except the first, has an initial fast drop (within $\approx2-3$\,s of the peak) followed by a slower exponential fade, lasting the remainder of the time (up to $\approx40$\,s) when the burst is observable over the persistent flux. This fast drop is by a greater factor in brighter bursts (Figure~\ref{fig:lc_stack}); for example, this drop is by a factor of $\approx2.5$ in burst~1 but $\approx4$ in burst~4.
This shape is typical of Type I bursts fuelled by helium \citep{galloway08rxte}.
Helium fuelled bursts can arise either when the accreted fuel is hydrogen poor or when accreted hydrogen burns stably between bursts; the binary orbital period is too long ($\approx76840$\,s, \citealt{buisson20atel2}) for a helium white dwarf companion and hydrogen is present in the optical spectra of the accretion disc/wind \citep{munoz20} so the latter case is more likely. The upper limits on the burst recurrence time (1.4\,days in the best case) are long enough that sufficient hydrogen burning is plausible.
The first burst is considerably fainter (peaking at $\approx290$\,cts/s over $0.5-10$\,keV; the next faintest, burst~4, peaks at $\approx600$\,cts/s) and shorter than the others.
Apart from the fourth, each burst is stronger than the previous one, while the persistent count rate decreases; this could be due to partial burning of the accreted material occurring outside bursts producing more H-poor fuel in the latter bursts, if more inter-burst burning occurred due to a longer recurrence time.

\subsection{Time resolved spectroscopy}
\label{sec:spec}
\begin{figure}
\includegraphics[width=\columnwidth]{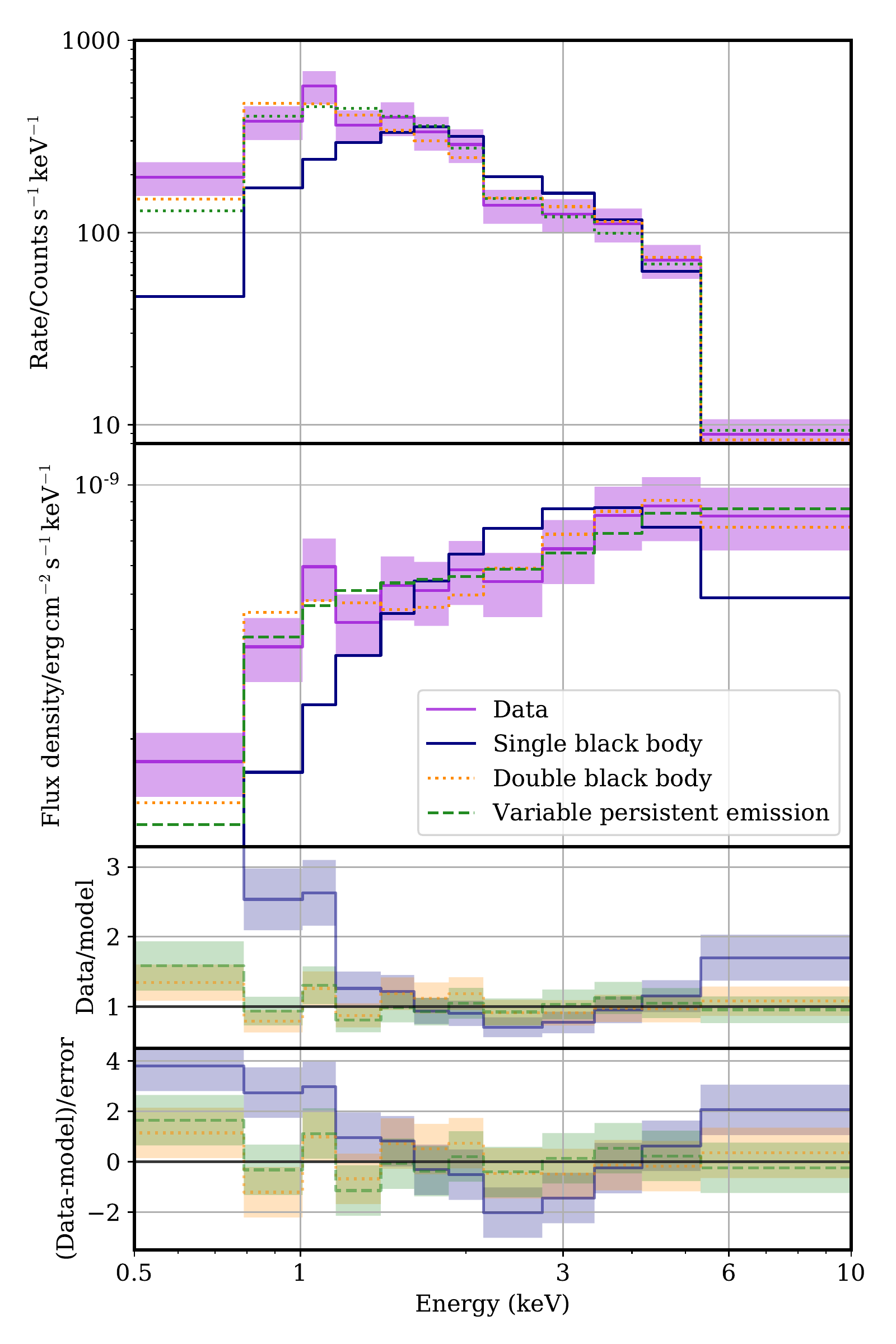}%
\caption{Comparison of different models for the net burst emission at the peak of the burst (the spectrum with highest count rate in Burst~5). A single blackbody (navy) is a poor fit; two blackbodies (yellow) or a contribution proportional to the persistent flux (green) both give similarly good fits.}
\label{fig:spec}
\end{figure}
\begin{figure*}
\includegraphics[width=\textwidth]{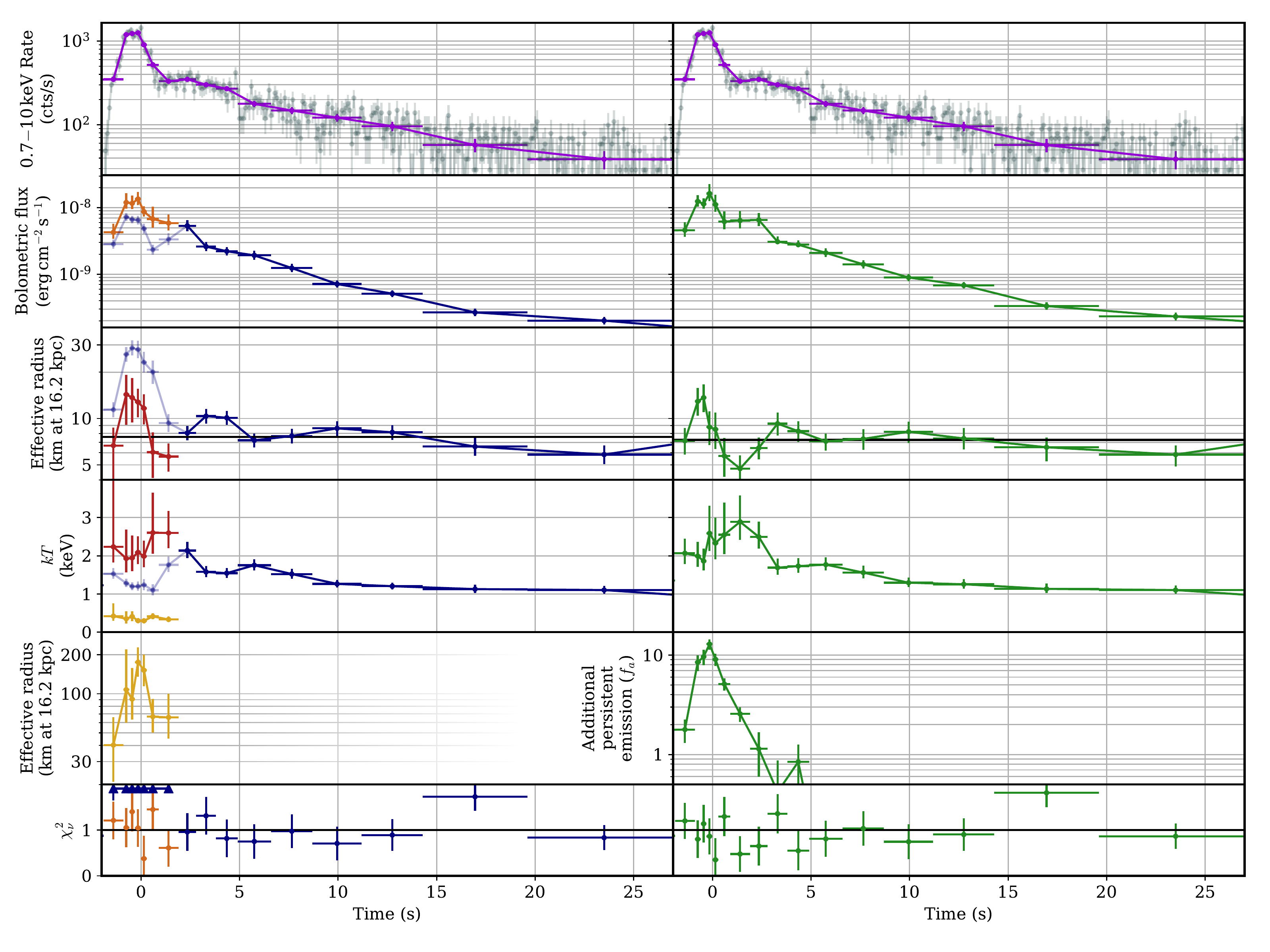}
\caption{Parameters of time-resolved spectra of Burst~5. Left column: Parameters from modelling the Type I burst emission with 1 or 2 blackbodies. Parameters for a single blackbody model are shown in navy; for two blackbodies, values for the complete model are shown in orange; the hotter blackbody in red and the cooler blackbody in yellow. The two-blackbody model is only shown where the single blackbody has $\chi^2_\nu>2$ and the single blackbody model is shown in faint navy in these cases.
Right column: Parameters from modelling the burst with a blackbody and an increase in the normalisation of the persistent emission (green).
Top panels: Lightcurves at 0.1\,s resolution (grey) and binned to the times of the spectra (purple).}
\label{fig:pars5}
\end{figure*}

We extract time-resolved spectra for each Type I burst using time intervals containing a minimum number of photons.
First, we estimate the persistent emission from the interval from 200\,s to 50\,s before the burst peak.
We then define the start of the burst: we take a light curve binned to 0.1\,s and find the final bin before the burst which is not above the persistent rate. We define the burst as starting at the end of this bin. Starting from this point, we extract spectra from time intervals containing at least 300 counts in excess of that expected from the persistent rate.
We then fit the spectrum of the burst emission as the difference between each burst spectrum and the persistent spectrum (this is performed by treating the persistent emission as the background).
We initially model the burst emission with a single blackbody.
Apart from around the burst peaks, the spectra are described well by this model. However, spectra around the peaks of the second, third and fifth bursts are broader than a simple blackbody and in the fits show excess emission at low  energies.
We test two alternative phenomenological models to explain this excess: allowing the normalisation of the persistent emission to change by a factor $(1+f_{\rm a})$ \citep{worpel13} or adding a second blackbody.
The former case requires a model for the persistent emission; we use \textsc{tbabs$\times$(diskbb+bbody)}, representing an absorbed disc and blackbody (we also use this model for the accretion rate estimates in Section~\ref{sec:distance}). This soft state model gives a good fit to the persistent spectra for each burst (the worst case $\chi^2/{\rm d.o.f.}=226.0/212=1.066,\ p=0.24$) and agrees with various properties of the bursts (mentioned throughout this work) which match other bursts observed during the soft state.

Both of these burst models provide good fits to all spectra (Figure~\ref{fig:spec}) and provide similar peak fluxes and qualitative behaviour of the first blackbody component's radius around the burst peak. The total fit statistics for burst~5 for the spectra from times where a single black body gives a poor fit ($\chi^2_\nu>2$) are: $\chi^2/{\rm d.o.f.}=78.7/76$ for the double black body model; and $\chi^2/{\rm d.o.f.}=73.5/83$ when varying the persistent emission. This implies a weak preference for a change in the strength of the persistent component but both models are statistically acceptable so we regard both options as possible.
Parameters of the fits of each of these models to burst 5 are shown in Figure~\ref{fig:pars5}.
Bursts 2 and 3 show similar features with lower signal; burst 1 has much lower signal; and we do not analyse burst 4 in detail due to the enhanced background from the SAA.

The area of the blackbody increases around the Type I burst peak before reducing and settling to a steady value for the majority of the burst tail, characteristic of photospheric radius expansion.
Both well-fitting models (two blackbodies or additional persistent emission) show a similar degree of expansion, by around a factor of 2 over the radius in the tail of the burst, once the radius has settled to a steady value.
The dip in blackbody radius after the burst peak to below the tail value is typical of bursts while accreting in the soft state, which agrees with our identification from the persistent spectrum, and is likely due to a changing colour-correction factor \citep{guver12i,guver12ii,kajava14}.

These fits show a fast rise and smooth decay in bolometric flux.
The apparent double peak in the flux curve for the single blackbody model is likely due to the poor fit around this time, although double peaks in bolometric luminosity have been seen in other PRE bursts \citep{jaisawal19}.
The comparatively smooth flux profile contrasts with the fast drop in count rate after the peak; the difference being due to the higher temperatures early in the decay producing a lower count rate for a given flux (when convolved with the instrument response, given the \nicer\ effective area curve and the temperatures concerned).

Near the times of the Type I burst peaks (within about 2\,s), there is an excess of soft emission over the simple black body model.
Similar excesses have been seen in Type~I bursts in many other sources observed with \nicer, e.g. Aql~X-1 \citep{keek18a}, 4U~1820--30 \citep{keek18b} and SAX~J1808.4--3658 \citep{bult19}.
This could be due to other extra components such as re-emission from the disc \citep[corresponding to the extra black body,][]{keek18a} or enhanced accretion through Poynting-Robertson drag \citep[corresponding to the change in persistent emission normalisation,][]{worpel13}. There may also be deviation from a simple blackbody due to Comptonisation \citep{keek18b} or scattering processes in the atmosphere \citep{romani87}.
The data for the X-ray bursts presented here are not sensitive enough to distinguish between these possibilities clearly.

\subsection{Distance estimate and implications}
\label{sec:distance}

\begin{figure}
\includegraphics[width=\columnwidth]{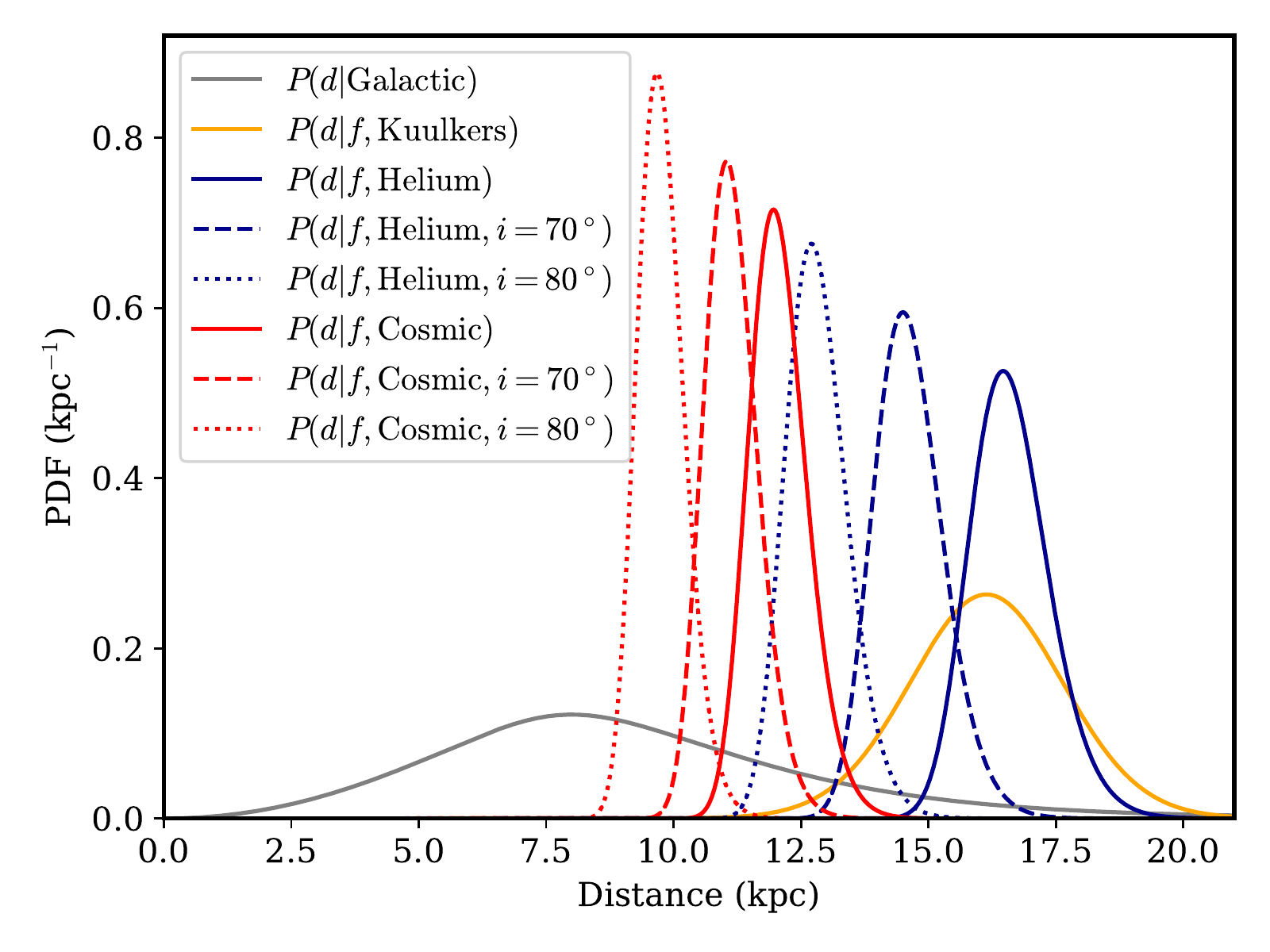}
\caption{Distance estimates for \eighteen\ based on various critical luminosities for PRE. Distance estimates for several specific luminosity values are shown (details in legend and text) along with the empirical luminosity range found by \citet{kuulkers03}. The Galactic prior is shown in grey.}
\label{fig:dist}
\end{figure}

Since the later Type I bursts (certainly burst 5, with some evidence also in bursts 2 and 3) show photospheric radius expansion, their peak luminosity should be governed by the Eddington limit. The observed flux can then be used to estimate a distance.
Initially, we use $L_{\rm Edd}=3.79\times10^{38}$\,erg/s, found empirically by \citet{kuulkers03} to be suitable for neutron stars at known distance, and to have an accuracy of 15\% for source-to-source variation. This matches the Eddington limit of a helium atmosphere around a $1.4M_\odot$ object.

We take the peak flux from the second, third and fifth Type I bursts (which are consistent with each other; different temperatures mean that these correspond to different count rates). We use the model of the burst including a scaled persistent emission component (see Section~\ref{sec:spec}), although the double black body model gives very similar results. For each burst, we use the least squares average of the fluxes from intervals which are consistent within 1-$\sigma$ of the highest value. These values are consistent with each other and their average is $1.1\pm0.1\times10^{-8}$\,erg\,cm$^{-2}$\,s$^{-1}$, which gives a distance of

$$D=16.7_{-1.3}^{+1.8}\,{\rm kpc}\ (1\sigma).$$

This puts \eighteen\ at the far side of the Galaxy; given its Sky coordinates ($l=26.3894$, $b=-5.3237$), this distance gives a Galactic (cylindrical) radius of $10$\,kpc and a height of $1.5$\,kpc below the Galactic plane.

Applying a prior for the relative density of the Galaxy along the line of sight (using the Galaxy model of \citealt{dehnen98,grimm02}; see also \citealt{gandhi19}) reduces this distance slightly, due to the higher density of objects closer to the Galactic centre, giving

$$D=16.2_{-1.6}^{+1.5}\,{\rm kpc}\ (1\sigma).$$

From the relative densities of the components of the Galactic model at this position, we infer that \eighteen\ is most likely (75\%) to be a disc object but could also be part of the halo (25\%). A bulge origin is highly unlikely ($P({\rm Bulge})=7\times10^{-6}$).

There are systematic effects which may affect this distance estimate \citep[e.g.][]{galloway08bias}.
Many of these, such as differences in neutron star mass and photosphere metallicity, are implicitly included by the empirical nature of the critical luminosity (and its uncertainty) measured by \citet{kuulkers03}.
However, the effects of obscuration in high inclination sources are not accounted for -- \citet{kuulkers03} find that in some high inclination sources the observed PRE luminosity is significantly lower. 
In this case, the photosphere may be partially obscured by larger components of the system, principally the disc.
For the simple case of a razor-thin disc, the disc can obscure up to half of the NS so the flux may be underestimated by up to a factor of 2 and the distance may actually be smaller by a factor of up to $\sqrt{2}$.
This factor is mitigated by reflection of the radiation intercepted by the disc but may be increased by a thick disc \citet{he16}.

\begin{table}
\caption{Distance estimates (kpc) for various gas compositions and inclinations (see text for details).}
\label{tab:dist}
\centering
\begin{tabular}{lccc}
\hline
 & Isotropic & $i=70^\circ$ & $i=80^\circ$\\
\hline
Pure helium & 
$16.6_{-0.8}^{+0.9}$&
$14.6_{-0.7}^{+0.8}$&
$12.8_{-0.6}^{+0.7}$\\
Cosmic abundances &
$12.1_{-0.6}^{+0.6}$&
$11.1_{-0.5}^{+0.6}$&
$9.8_{-0.4}^{+0.5}$\\
\hline
\end{tabular}
\end{table}

To show the magnitude of these effects, we show distance estimates for various specific values of metallicity and inclination in Figure~\ref{fig:dist} and Table~\ref{tab:dist}. We calculate the distances by replacing the empirical peak luminosity from \citet{kuulkers03} with the theoretical Eddington luminosity \citep[e.g.][]{lewin93} modified by the anisotropy factor ($\xi_{\rm b}$) from \citep{he16},
$$L_{\rm Obs}=\frac{8\pi Gm_{\rm p}M_{\rm NS}c}{\xi_{\rm b}\sigma_{\rm T}(1+X)(1+z(R))}$$
where $G$ is the gravitational constant, $m_{\rm p}$ is the proton mass, $M_{\rm NS}$ is the neutron star mass, $c$ is the speed of light, $\sigma_{\rm T}$ is the Thomson cross section, $X$ is the hydrogen mass fraction and $z(R)$ is the gravitational redshift at the photospheric radius $R$. We show the two extremes of likely metallicity, pure helium ($X=0$) and a cosmic abundance of hydrogen ($X=0.739$).
Since it shows eclipses \citep{buisson20atel2}, \eighteen\ is at high inclination; from \citet{he16}, the appropriate reduction in apparent luminosity ($\xi_{\rm b}^{-1}$) for inclinations of $70-80^\circ$ is a factor of $0.85-0.65$.
For each combination of parameters, we show the distance estimate for a neutron star mass of $1.4M_\odot$ and a photospheric radius of 20\,km.
This allows a significantly larger range of distances than the \citet{kuulkers03} range, due to the lower effective Eddington luminosities for high hydrogen fractions and high inclination. However, even the smallest distance estimate ($9.8_{-0.4}^{+0.5}$\,kpc) is 60\% of the value derived from the \citet{kuulkers03} luminosity and beyond the average distance of Galactic sources along this line of sight.

To reduce the range of these estimates, we can consider whether particular values of parameters generating the systematic uncertainty are preferred by other evidence.
The eclipse 	duration implies an inclination of at least 70$^\circ$ \citep[Buisson et al. in prep.]{buisson20atel2}. A more accurate determination of the inclination would require detailed modelling of optical light curves and spectra, beyond the scope of this paper; meanwhile, we regard our calculation using 80$^\circ$ as a fiducial value.
The atmospheric composition of a Type I burst can be inferred from its light curve. The relatively fast rise and initial decay of the PRE bursts observed here suggest a helium burst. Additionally, helium fuelled bursts are also more common during the soft accretion state and the dip in apparent radius below the final value is more typical of soft state bursts \citep{kajava14}.
Further, bursts can reach the Eddington limit for helium even where accreted material is hydrogen rich, either by the hydrogen being burnt between bursts or the hydrogen rich atmosphere being blown off by the burst \citep{bult19,galloway06}.
 This would imply that the further distance estimates (blue curves in Figure~\ref{fig:dist}) are more likely ($12.8_{-0.6}^{+0.7}$\,kpc for $i=80^\circ$).

With this distance estimate, we can also estimate the accretion rate at the times of the bursts from the persistent flux measurements from the modelling in Section~\ref{sec:distance}. We find a bolometric (of the X-ray components) flux before the bursts of $12.6_{-0.4}^{+0.5}$, $10.2_{-0.4}^{+0.5}$, $8.9_{-0.4}^{+0.6}$, $5.6_{-0.3}^{+0.5}$ and $2.3_{-0.15}^{+0.2}$ $\times10^{-10}$ erg\,cm$^{-2}$\,s$^{-1}$, in chronological order.
If the persistent emission has the same anisotropy as the burst, this implies an Eddington fraction $\dot{m}_{\rm Edd}=0.20_{-0.02}^{+0.01}, 0.16_{-0.02}^{+0.01}, 0.14_{-0.02}^{+0.01}, 0.09_{-0.01}^{+0.01}$ and $0.036\pm0.004$ for material in the accretion flow (calculating $L_{\rm Edd}$ for $X=0.73$). However, the disc and boundary layer may have more anisotropic emission than the burst from the NS surface \citep[e.g.][]{he16}, so the true Eddington fraction could be somewhat higher. The exact factor depends on the details of the accretion structure and the inclination; for a flat disc (which provides all the persistent flux) observed at 70-80$^\circ$, the increase is by a factor of 1.2-2.
This is similar to the range at which helium fuelled bursts are expected and observed \citep{galloway08rxte} but extends slightly higher, so there could be some influence of residual hydrogen in the burning material.

Our distance estimates are all relatively large \citep{galloway08rxte,gandhi19} but not unprecedented \citep[e.g.][]{homan14} for an XRB.
The absorbing column density ($\approx2\times10^{21}$\,cm$^{-2}$) is comparatively low for such a distant source, but the total Galactic column density in the direction of \eighteen\ is similar \citep[$1.8\times10^{21}$\,cm$^{-2}$,][]{hi4pi16}.

A large distance can also help in explaining the strong variability observed in the initial state of \eighteen\ (during 2018-9): it is comparatively faint for a binary but strong winds \citep{munoz20} and variability \citep{ludlam18_1858} are often explained by a high Eddington rate \citep{king03,grupe04}.
During the flaring state but between flares, the observed flux of \eighteen\ was $\approx2.5\times10^{-10}$\,erg\,cm$^{-2}$\,s$^{-1}$ \citep{hare20}\footnote{This was measured for the 3-78\,keV band, which for the spectral shape of this observation includes the majority of flux; any bolometric correction will only increase the strength of super-Eddington behaviour.}, which is $\approx5$\% of the Eddington limit for hydrogen and a $1.4M_\odot$ object.
Some flares increased count rates by factors of many tens, so at least during bright flares, the luminosity was above the Eddington luminosity (and correcting for any anisotropy is likely only to increase the strength of this). If much of the variability was due to obscuration, the intrinsic luminosity would also have been above Eddington at other times.

\subsection{Pre-burst oscillations}

\begin{figure}
\includegraphics[width=\columnwidth]{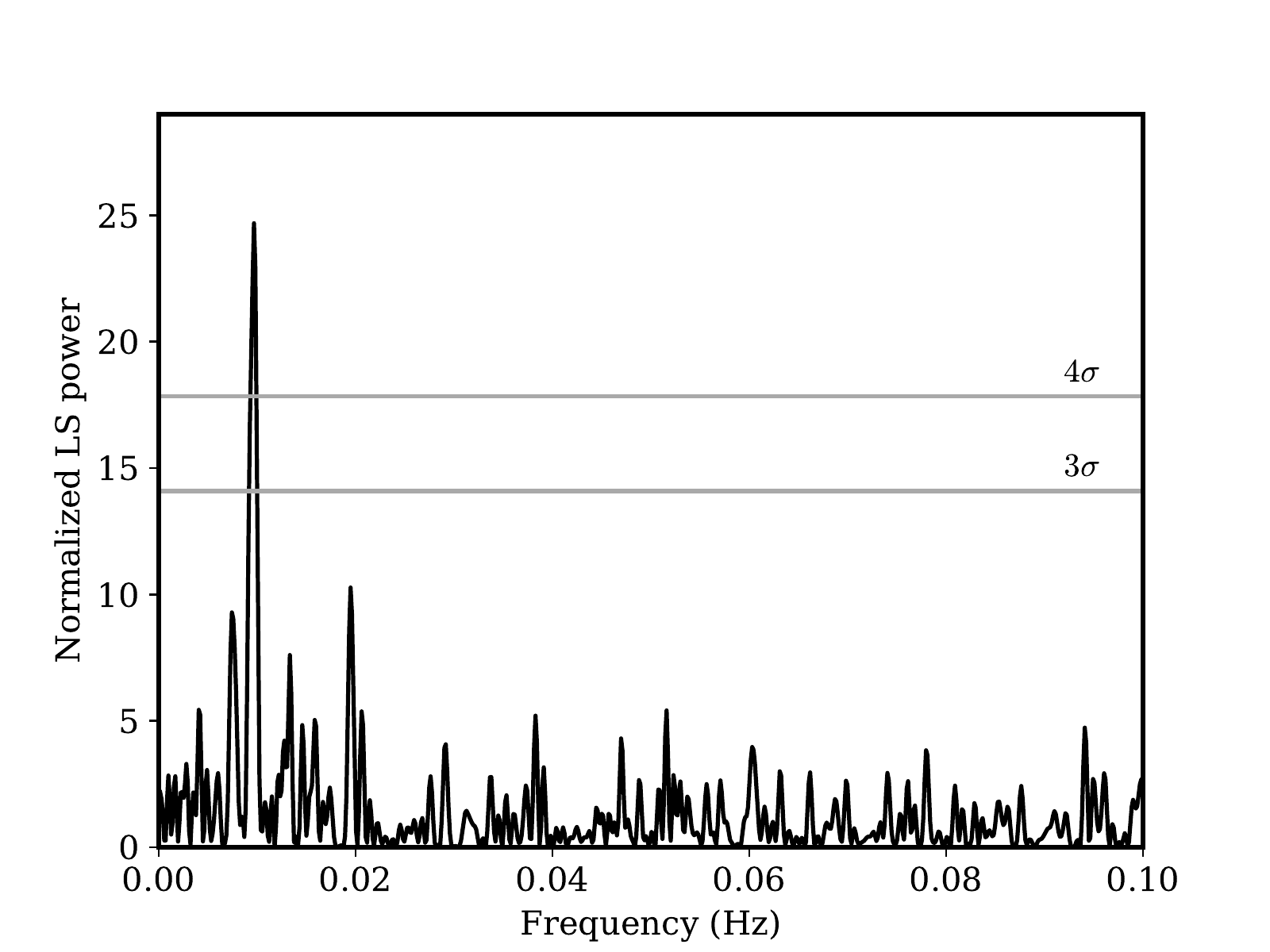}
\includegraphics[width=\columnwidth]{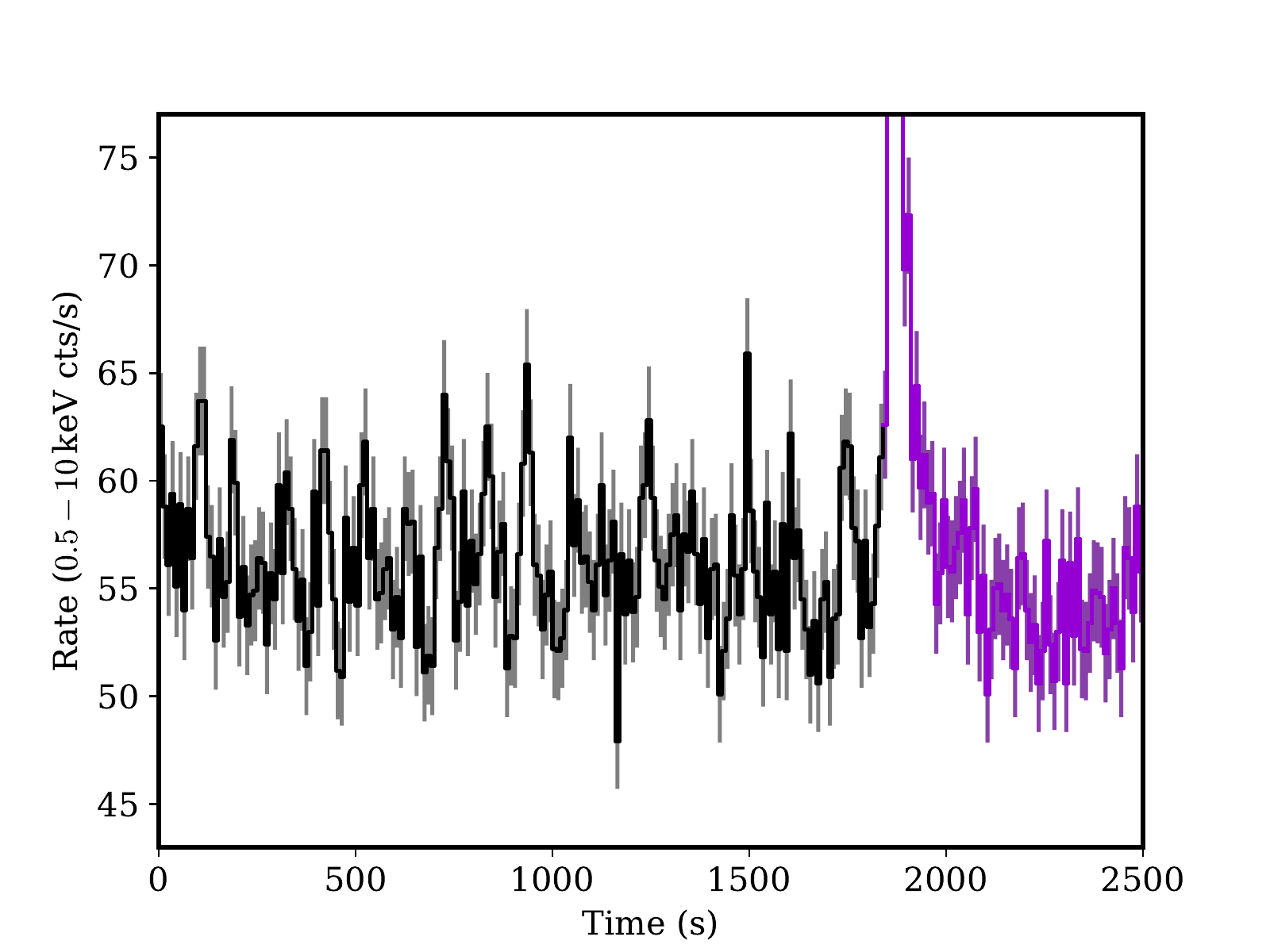}
\caption{Top: Lomb-Scargle periodogram of the light curve segment immediately prior to burst 5, which shows a QPO at $9.6\pm0.5$\,mHz.
Bottom: $0.5-10$\,keV \nicer\ light curve of the period up to and including burst 5, during which a $9.6\pm0.5$ mHz QPO is detected. The burst itself (purple) far exceeds the plotted range.}
\label{fig:qpo}
\end{figure}

We also looked for millihertz quasi-periodic oscillations (mHz QPOs), which are sometimes found before an X-ray burst \citep[e.g.][]{revnivtsev01,altamirano08,mancuso19}. We used $0.5-10$\,keV light curves at 1\,second resolution and applied the Lomb-Scargle periodogram \citep{lomb76,scargle82} to each gap-less light curve, excluding periods of dipping and eclipses. In the five cases where we detected the type-I X-ray bursts, we searched for the oscillations before and after the X-ray bursts. To estimate the significance level, we followed the approach of \citep{press92}, which assumes white noise and takes as a number of trials the number of frequencies explored.

We detected a mHz QPO at a significance of $5.8\sigma$ in the 1.8\,ks of data before the 5th X-ray burst (Figure~\ref{fig:qpo}). The mHz QPO has an average frequency of $9.6\pm0.5$\,mHz and a fractional rms amplitude of $2.2\pm0.2$\% ($0.5-10$\,keV). There is no evidence of the oscillations in the $\approx600$\,s of data after the X-ray burst, with a 90\% upper limit on the rms amplitude of 1.2\% ruling out that the same strength of oscillation continues.
We also found marginal evidence of QPOs in at least three other cases; however the datasets are relatively short ($\lesssim 500-700$\,s), and therefore it is not possible to understand if they are real or the product of red-noise. 
The upper limits to the QPO amplitude for the time segments prior to the earlier bursts are somewhat lower in fractional amplitude (0.6, 0.8, 1.0 and 0.8\% in chronological order) than for the detected QPO but due to the brighter flux at earlier times, all but the last of these limits are higher in absolute amplitude than the detected QPO. Therefore, we cannot definitively rule out a QPO from the same level of oscillatory burning occurring prior to the other bursts.
We also note that a mHz QPO was also detected in a \nustar\ observation during 2019~February \citep{hare19}, although at a frequency of 2.7\,mHz, which is lower than other mHz QPOs which have been explained by marginally stable nuclear burning.

\citet{revnivtsev01} find that these mHz QPOs are only found in a narrow range of luminosities, $L_{\rm 3-20\,{\rm keV}}=5-11\times10^{36}$\,erg\,s$^{-1}$. 
The QPO found here occurs while the mean flux (from our model for the persistent emission, Section~\ref{sec:spec}) is $f_{\rm 3-20\,{\rm keV}}=1.4\pm0.2\times10^{10}$\,erg\,cm$^{-2}$\,s$^{-1}$, which corresponds to $L_{\rm 3-20\,{\rm keV}}=4.3_{-0.5}^{+0.6}\times10^{36}$\,erg\,s$^{-1}$ at $16.2$\,kpc (derived from the Type~I burst peak luminosity for the population, since we are comparing with a population luminosity for other QPO detections), which supports the upper end of the distance estimates found here. Using a lower distance to account for the effect of high inclination in \eighteen\ would imply a lower luminosity, which is compatible with the luminosities at which QPOs are observed in other sources if this also depends on inclination, with at least as strong an anisotropy factor as the burst emission. This would be expected for disc, NS surface or coronal emission so long as this is not strongly equatorially beamed.

The characteristics of the mHz QPOs we found here are consistent with those found in 6 other NS systems \citep[but compare \citealt{linares12}]{revnivtsev01,altamirano08,strohmayer11,lyu14,lyu15,strohmayer18,mancuso19} and are usually explained as being the product of marginally stable burning of He on the NS surface \citep{heger07qpo}. This is the 7$^{\rm th}$ NS system that shows this type of QPOs. The fact that we do not detect more episodes of mHz QPOs could either be due to their intrinsic absence or to detection difficulty: the mHz QPOs are not always present in the X-ray light curves (they are state dependent and even in a given state, there is not yet a clear physical trigger for them, see \citealt{altamirano08,mancuso19}, etc.); in addition, the frequency and amplitude of these QPOs are very low, and therefore to acquire enough QPO cycles and sufficient signal-to-noise, uninterrupted datasets longer than $1000-1500$\,s are generally needed to unambiguously detect them. 

\subsection{Burst oscillation search}

We searched each of the X-ray bursts observed with {\it NICER} for the presence of burst oscillations,
but did not detect any significant signals. To search for oscillations, we constructed a 1/8192\,s time resolution light 
curve for each X-ray burst, using only those events in the $1-8$\,keV energy band. These light curves all started 10 
seconds prior to the burst onset, and had durations of 40 seconds. For each considered X-ray burst, we applied a 
$T=2,4,8$\,s duration window selection, which we moved across the burst profile in steps of $T/2$. We then calculated 
the power spectrum associated with each window position, and searched the $100-2000$\,Hz frequency range for 
excess power over the expected noise distribution. No such excess was observed, to a 95\% confidence upper limit 
of approximately 15\% fractional amplitude in the most sensitive segment (the peak light curve of burst 5). We note, 
however, that the vast majority of considered segments had much lower averaged count-rates, and thus substantially 
higher upper-limits. With typical upper limits ranging between 30\% and 80\% fractional amplitude, our results are 
therefore not especially constraining. 

\section{Further discussion}
\label{sec:dis}

\subsection{Implications of the neutron star accretor}

The identification of the accretor in \eighteen\ as a neutron star informs several outstanding questions relating to the properies of \eighteen.
It fits with the low coronal temperature found in \citet{hare20}, since neutron stars tend to have lower coronal temperatures than black holes \citep{burke17}.
 
However, the neutron star accretor implies an unusual location in the radio-X-ray plane: \eighteen\ appears relatively X-ray faint for a NS XRB \citep{vandeneijnden20}.
This could imply that \eighteen\ has an intrinsically unusually low X-ray/radio luminosity ratio or that the observed X-ray luminosity is unrepresentatively low. The latter case would support a model in which the X-ray emission (which may already be comparatively low due to anisotropy, e..g. \citealt{he16}) is usually obscured by the high inclination disc, apart from during the flares, which  represent the true intrinsic luminosity, when viewing the central source directly through a gap in the (irregular) disc surface.

\eighteen\ has previously been compared with the black hole XRBs \vsgr\ and \vcyg\ \citep[e.g.][]{hare20}. All of these sources have shown strong variability due to some combination of changes in intrinsic flux and obscuration, although the relative contribution of these two effects is not yet clear (e.g. compare \citealt{walton17,koljonen20}).
The relative radio loudness also provides a further similarity with \vcyg, which is unusually radio loud for its inclination \citep{motta18}.
The identification of \eighteen\ as a neutron star XRB means the flaring behaviour in these sources must now be explained in a model which is compatible with a neutron star accretor.
In particular, extreme variability from processes very close to the event horizon may be ruled out, since a neutron star is significantly larger than its Schwarzschild radius.

\subsection{Bursts in the flaring state?}

\eighteen\ had been active for over a year before any Type I bursts were detected; there are several means to explain the non-detection of bursts during this period.
Firstly, there may truly have been no bursts, due to the different accretion regime during this period.
In a model where variable obscuration causes much of the strong variability, the intrinsic accretion rate was much higher during the flaring period, so would likely have induced stable nuclear burning of both hydrogen and helium.
Additionally, in this model, the obscuration between flares would have impeded observation of any Type~I bursts which occurred while the neutron star was obscured (which is the majority of the duty cycle).
It is also possible that bursts were observed but not identified if they occurred at the same time as flares.
The observed flares are all different in spectrum, light curve and/or duration to thermonuclear bursts; however, the variety of flares means that it is possible that a burst coincident with a flare would go unnoticed.
Finally, it is also possible that bursts did occur during this phase of the outburst but, by chance, not during \nicer\ observations of \eighteen.
Overall, it is unsurprising that X-ray bursts had not been detected in the flaring state, whether or not they occurred.

\subsection{Comparison with other similar sources}

We can also compare the flaring state to other strong variability regimes in neutron stars.
Two famous neutron star systems exhibiting flare-like behaviour are the Rapid Burster \citep[MXB~1730--335, e.g.][]{hoffman78} and Bursting Pulsar (GRO~J1744--28, \citealt{fishman95}). The Rapid Burster shows many (up to thousands per day) `rapid' bursts in addition to Type I bursts; these rapid bursts are much shorter ($<10$\,s) and more regular in cadence than the flares of \eighteen, so are probably different phenomena.
The Bursting Pulsar is the archetypal example of Type~II X-ray bursts \citep{kouveliotou96}. These bursts also differ markedly from the flares observed in \eighteen: the type~II bursts are again much shorter and are accompanied by a drop in emission following the burst.
Therefore, the flaring state of \eighteen\ is not explained as an example of these other unusual neutron star XRB states.

The high inclination NS LMXB EXO~0748--676 has also shown flaring episodes \citep{homan03}, although these are more sporadically interspersed with other light curve shapes and less prominent at harder energies than those in \eighteen.

Transitional millisecond pulsars (tMSPs) also have a `flaring' accretion mode \citep{demartino13,bogdanov15}, although this occurs at much lower luminosity ($\approx10^{34}$\,erg\,s$^{-1}$) than the flaring state in \eighteen\ ($\gtrsim10^{36}$\,erg\,s$^{-1}$ observed). The tMSP flaring mode can also show strong, variable absorption \citep[e.g.][]{li20}, so could be an analogue with lower accretion efficiency.

There have not yet been measurements of the magnetic field strength in \eighteen; the closer comparison of the flaring state of \eighteen\ with black hole than neutron star systems could be because the magnetic field of its neutron star is low enough to be unimportant in its accretion flow, implying a relatively low magnetic field strength.

\section*{Acknowledgements}

We thank Laurens Keek for helpful discussions and the referee for comments which improved the manuscript.
We thank the NuSTAR operations team for rapid approval and execution of our Target of Opportunity proposal.
D.J.K.B. and D.A. are funded by the Royal Society.
T.G. has been supported in part by the Scientific and Technological Research Council (T\"UBITAK) 119F082, Royal Society Newton Advanced Fellowship, NAF$\backslash$R2$\backslash$180592, and Turkish Republic, Directorate of Presidential Strategy and Budget project, 2016K121370.
J. Hare acknowledges support from an appointment to the NASA Postdoctoral Program at the Goddard Space Flight Center, administered by the Universities Space Research Association under contract with NASA.
C.M. is supported by an appointment to the NASA Postdoctoral Program at the Marshall Space Flight Center, administered by the Universities Space Research Association under contract with NASA.
M.O.A. acknowledges support from the Royal Society through the Newton International Fellowship programme.
This work made use of data from the \nustar\ mission, a project led by the California Institute of Technology, managed by the Jet Propulsion Laboratory, and funded by the National Aeronautics and Space Administration. This research has made use of the \nustar\ Data Analysis Software (NuSTARDAS) jointly developed by the ASI Science Data Center (ASDC, Italy) and the California Institute of Technology (USA).
\nicer\ is a mission of NASA's Astrophysics Explorers Program. This research has made use of data and software provided by the High Energy Astrophysics Science Archive Research Center (HEASARC), which is a service of the Astrophysics Science Division at NASA/GSFC and the High Energy Astrophysics Division of the Smithsonian Astrophysical Observatory.
This research has made use of ISIS functions (ISISscripts) provided by ECAP/Remeis observatory and MIT (http://www.sternwarte.uni-erlangen.de/isis/).

\section*{Data Availability}

The data underlying this article are available in HEASARC.

\bibliographystyle{mnras}
\bibliography{swiftj1858}

\appendix

\section{Summary of burst properties}

\begin{table*}
\caption{Summary of burst properties.}
\label{tab:summ}
\rotatebox{-90}{
\centering
\begin{tabular}{lcr@{}lr@{}lr@{}lcr@{}l}
\hline
Burst & Recurrence time$^{\rm a}$ & \multicolumn{2}{c}{Peak rate$^{\rm b}$} & \multicolumn{2}{c}{Persistent flux$^{\rm c}$} & \multicolumn{2}{c}{Persistent accretion rate$^{\rm d}$} & mHz QPO frequency$^{\rm e}$ & \multicolumn{2}{c}{mHz QPO amplitude$^{\rm e}$} \\
number  & (days) & \multicolumn{2}{c}{($0.7-10$\,keV cts/s)} & \multicolumn{2}{c}{($\times10^{-10}$ erg\,cm$^{-2}$\,s$^{-1}$)} & \multicolumn{2}{c}{($\dot{M}_{\rm Edd}$)} & (mHz) & &\\
\hline
1 &       $-$ & \ \ \ \ \ \ \ \ 550&$\pm$75& \ \ \ \ \ \ \ $12.$&$6_{-0.4}^{+0.5}$  &  \ \ \ \ \ \ \ \ \ \ $0.$&$20_{-0.02}^{+0.01}$ &    $-$      &  \ \ \ \ \ \ \ \ \ \ $<0.$&$6$\% \\
2 & $\leq4.5$ & 960&$\pm$100  & $10.$&$2_{-0.4}^{+0.5}$  & $0.$&$16_{-0.02}^{+0.01}$ &    $-$      & $<0.$&$8$\% \\
3 & $\leq3.6$ & 1190&$\pm$110 &  $8.$&$9_{-0.4}^{+0.6}$  & $0.$&$14_{-0.02}^{+0.01}$ &    $-$      & $<1.$&$0$\% \\
4 & $\leq6.3$ & 740&$\pm$90   &  $5.$&$6_{-0.3}^{+0.5}$  & $0.$&$09_{-0.01}^{+0.01}$ &    $-$      & $<0.$&$8$\% \\
5 & $\leq1.4$ & 1490&$\pm$120 &  $2.$&$3_{-0.15}^{+0.2}$ & $0.$&$036\pm0.004$ & $9.6\pm0.5$ & $2.$&$2\pm0.2$\% \\
\hline
\end{tabular}
}

$^{\rm a}$Recurrence times are upper limits as bursts may have occurred between observations. $^{\rm b}$Peak rate is the highest value in the $0.7-10$\,keV light curve binned to 0.1\,s. $^{\rm c}$Persistent flux is measured bolometrically. $^{\rm d}$The persistent accretion rate is not corrected for the effects of anisotropy, which likely increase it by a factor of 1.2-2. $^{\rm e}$The mHz QPO is measured over 0.5-10\,keV.

\end{table*}

\clearpage
\section*{Affiliations}
  $^{1}$Department of Physics and Astronomy, University of Southampton, Highfield, Southampton, SO17 1BJ\\
  $^{2}$Department of Astronomy, University of Maryland, College Park, MD 20742, USA\\
  $^{3}$NASA/Goddard Space Flight Center, Code 662, Greenbelt, MD 20771, USA\\
  $^{4}$Instituto Argentino de Radioastronom\'{i}a(CCT-La Plata, CONICET; CICPBA), C.C. No. 5, 1894 Villa Elisa, Argentina\\
  $^{5}$Facultad de Ciencias Astron\'{o}micas y Geof\'{i}sicas, Universidad Nacional de La Plata, Paseo del Bosque s/n, 1900 La Plata, Argentina\\
  $^{6}$Istanbul University, Science Faculty, Department of Astronomy and Space Sciences, Beyaz\i t, 34119, Istanbul, Turkey\\
  $^{7}$Istanbul University Observatory Research and Application Center, Istanbul University 34119, Istanbul Turkey\\
  $^{8}$National Space Institute, Technical University of Denmark, Elektrovej 327-328, DK-2800 Lyngby, Denmark\\
  $^{9}$MIT Kavli Institute for Astrophysics and Space Research, Massachusetts Institute of Technology, Cambridge, MA 02139, USA\\
  $^{10}$IRAP, CNRS, UPS, CNES, 9 avenue du Colonel Roche, BP 44346, F-31028 Toulouse Cedex 4, France\\
  $^{11}$Eureka Scientific, Inc., 2452 Delmer Street, Oakland, CA 94602, USA\\
  $^{12}$SRON, Netherlands Institute for Space Research, Sorbonnelaan 2, 3584 CA Utrecht, The Netherlands\\
  $^{13}$Department of Astronomy, Tsinghua University, Shuangqing Road 30, Beijing 100084 China\\
  $^{14}$Tsinghua Center for Astrophysics, Tsinghua University, Shuangqing Road 30, Beijing 100084 China\\
  $^{15}$NASA Marshall Space Flight Center, NSSTC, 320 Sparkman Drive, Huntsville, AL 35805, USA\\
  $^{16}$Universities Space Research Association, Science and Technology Institute, 320 Sparkman Drive, Huntsville, AL 35805, USA\\
  $^{17}$Department of Astronomy, University of Michigan, 1085 South University Avenue, Ann Arbor, MI 48109-1104, USA\\
  $^{18}$Department of Astronomy \& Astrophysics, Atat\"{u}rk University, Erzurum, Turkey\\
  $^{19}$Astrophysics Science Division and Joint Space-Science Institute, NASA's Goddard Space Flight Center, Greenbelt, MD 20771, USA\\
  $^{20}$Department of Physics, Tor Vergata University of Rome, Via della Ricerca Scientifica 1, I-00133 Rome, Italy\\
  $^{21}$INAF - Astronomical Observatory of Rome, Via Frascati 33, I-00078 Monte Porzio Catone (Rome), Italy\\
  $^{22}$Space Sciences Laboratory, 7 Gauss Way, University of California, Berkeley, CA 94720-7450, USA\\
  $^{23}$Institute of Astronomy, Madingley Road, Cambridge, CB3 0HA\\

\bsp	
\label{lastpage}

\end{document}